\documentclass[aps,prl,groupedaddress,superscriptaddress,twocolumn,10pt,longbibliography,floatfix]{revtex4-2}
\pdfoutput=1 
\usepackage[utf8]{inputenc}
\usepackage{hyperref}
\usepackage{graphicx}
\usepackage{bm}
\usepackage{dsfont}
\usepackage{orcidlink}
\usepackage{comment}
\usepackage{booktabs}
\usepackage{tabularx}
\usepackage{cancel}
\usepackage{siunitx}

\newcommand{\cmtout}[1]{}

\usepackage{siunitx} 
\newcommand{\omatt}[1]{}

\newcommand{\identity}{\mathds{1}}
\newcommand{\discuss}[1]{}

\usepackage{soul} 
\usepackage[normalem]{ulem}
\usepackage{amssymb}
\usepackage{physics}
\usepackage{hyperref}
\DeclareSIUnit\bohr{\text{\ensuremath{a_\textup{0}}}}

\begin{document}
\title{Electron lattice potentials for ultracold atoms using circular Rydberg orbitals}
\date{\today}

\author{Aileen A. T. Durst\orcidlink{0009-0004-0498-5238}}
\affiliation{Max-Planck Institut für Physik komplexer Systeme, Nöthnitzer Stra\ss e 38, 01187 Dresden, Germany}
\affiliation{Joint Quantum Centre (JQC) Durham-Newcastle, Department of Chemistry,
Durham University, South Road, Durham, DH1 3LE, United Kingdom.}

\author{Einius Pultinevicius\orcidlink{0009-0005-7404-9178}}
\affiliation{5. Physikalisches Institut and Center for Integrated Quantum Science and Technology, Universität Stuttgart, Pfaffenwaldring 57, 70569 Stuttgart, Germany}

\author{Homar Rivera-Rodr\'iguez\orcidlink{0009-0007-6454-7063}}
\affiliation{Max-Planck Institut für Physik komplexer Systeme, Nöthnitzer Stra\ss e 38, 01187 Dresden, Germany}

\author{Tilman Pfau\orcidlink{0000-0003-3272-3468}}
\affiliation{5. Physikalisches Institut and Center for Integrated Quantum Science and Technology, Universität Stuttgart, Pfaffenwaldring 57, 70569 Stuttgart, Germany}

\author{Matthew T. Eiles\orcidlink{0000-0002-0569-7551}}
\affiliation{Department of Physics and Astronomy, Purdue University, West Lafayette, IN 47907, USA}

\author{Florian Meinert\orcidlink{0000-0002-9106-3001}}
\affiliation{5. Physikalisches Institut and Center for Integrated Quantum Science and Technology, Universität Stuttgart, Pfaffenwaldring 57, 70569 Stuttgart, Germany}

\begin{abstract}
Recent advances in experiments using individually trapped atoms have enabled precise control over circular Rydberg electrons with exceptionally long lifetimes. We show that these giant and stable electron orbits can form toroidal lattice potentials for ultracold atoms with a period set by the electron's de Broglie wavelength. Unlike conventional static optical lattices, this electron lattice is formed via the electron-atom interaction, which mixes Rydberg circular states with opposite azimuthal phase winding into a standing electronic matter  wave. The lattice phase is thereby intrinsically coupled to the atom position. For a pair of atoms, this results in ballistic tunneling motion in the two-atom spatial correlations along the ring, while the single particle dynamics is essentially free rotation. We simulate the dynamics for an experimentally realistic setting that exploits optical tweezers for individual atom control. Our results open a route toward incorporating long range atom-atom interactions mediated by a single electron and, ultimately, realizing small Bose and Fermi gases confined in these microscopic electronic atom traps. 
\end{abstract}

\maketitle

Potential landscapes generated by electronic matter waves are a well-established concept in condensed matter physics. Prominent examples include charge-density waves, which can induce distortions of the underlying ionic lattice \cite{Gruner1988}, and polarons, which emerge from the coupling of the electronic density to lattice phonons \cite{Landau1933}. Ultracold atoms confined in optical lattices provide a versatile platform for simulating the behavior of strongly interacting electrons in solids \cite{Gross2017,Greiner2002,Jördens2008,Mazurenko2017}. However, engineering a dynamical backaction of the trapped atoms onto the imposed static lattice potential remains an outstanding challenge \cite{Baumann2010,DeSalvo2019}. 

Here, we propose and theoretically investigate a novel approach for realizing an electronic lattice that forms due to the presence of ultracold atoms, which, in turn, propagate through the lattice. In our scheme, the trapping potential is generated by the matter wave of a weakly bound electron prepared in a circular Rydberg state. \cite{Hulet1983,Hoelzl2024,Cortinas2020}.
Owing to their large orbital angular momentum, these states exhibit exceptionally long lifetimes. Recent experiments with individually trapped circular Rydberg atoms have demonstrated lifetimes exceeding 10 ms for principal quantum numbers around ($n\approx100$) \cite{Pultinevicius2025}. 
This stability provides ample opportunity to immerse and trap one or several ground-state atoms within the micrometer-sized Rydberg orbital and investigate their collective quantum dynamics.

\begin{figure}[!t]
  \centering
\includegraphics[width=\linewidth]{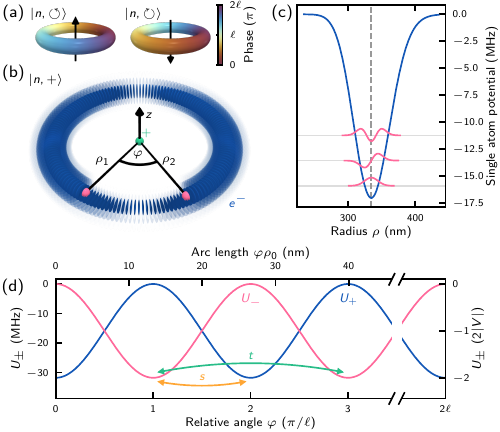}
  \caption{Electron lattice emerging from circular Rydberg states. (a) Ring-shaped orbitals of the $n=80$, $\ell = 79$ circular states with opposite magnetic quantum number $m=\pm\ell$. The shading encodes the unwrapped azimuthal phase $m\phi$. (b) Density of the electronic standing wave (blue) produced by a superposition of these states. The positions of two atoms (pink) sitting inside the circular orbit with respect to the Rydberg core (green) and their relative azimuthal coordinate $\varphi$ are indicated. (c) Radial cut through the Born-Oppenheimer potential energy surface at $z=0$ for a single atom. The energy of the first three radial vibrational states are depicted by horizontal lines. The vertical dashed line shows the equilibrium position at $\tilde{\rho}=n(n-1)a_0$, where $a_0$ is the Bohr radius. (d) Azimuthal electronic lattice potentials $U_\pm$ dictating the relative motion of two atoms inside the circular orbit. Arrows indicate tunneling processes within and between the sublattices, with $s = J\ell K g$ as discussed in the text. }
  \label{Fig1}
\end{figure}

The Fermi contact interaction between ground-state atoms and the Rydberg electron is known to generate binding potentials for ultralong-range Rydberg molecules \cite{Greene2000,Bendkowsky2009}. While these molecules have been extensively studied for short-lived, low-angular-momentum Rydberg states \cite{Shaffer2018,Fey2020,Dunning2024}, the electronic potentials created by circular Rydberg states remain largely unexplored. Within the hydrogenic manifold of Rydberg states with principal quantum number $n$, one finds two circular orbitals, $\lvert n,\circlearrowright\rangle=\lvert n,\ell=n-1,m=+\ell\rangle$, and $\lvert n, \circlearrowleft\rangle=\lvert n,\ell=n-1,m=-\ell\rangle$, where $\ell$ and $m$ are the orbital angular momentum and magnetic quantum numbers. Both of these states form a ring-shaped orbital but, because of the azimuthal phase factor $e^{-i m \phi}$, have opposite phase winding. As illustrated in Fig.~\ref{Fig1}(a,b), a superposition of the two states thus realizes an electronic standing wave along the ring with $2\ell$ density maxima separated by a distance $d\approx \pi a_0 n = 13.5\,\rm{nm}$ for $n=80$, where $a_0$ is the Bohr radius.

In the following, we show how this electron density generates MHz-deep toroidal lattice potentials governing the dynamics of the ground-state atoms, taken here to be Sr atoms. In contrast to conventional optical lattices \cite{Gross2017}, our electronic lattice is intrinsically coupled to the atomic motion through the underlying electron wavefunction. We identify experimentally accessible parameter regimes for realizing such an atomic-scale electronic lattice and propose an optical-tweezer-based protocol to observe correlated tunneling dynamics within this lattice.

A single ground-state atom inside the circular Rydberg orbit interacts with the Rydberg electron via Fermi's contact potential, $V_{\mathrm{ea}}(\bm r,\bm R) = \frac{2\pi a_s\hbar^2}{m_e}\delta^3(\bm r-\bm R)$, where $\bm r$ ($\bm R$) denotes the electron (atom) position and $a_s$ is the zero-energy electron-atom scattering length \cite{Greene2000,Fermi1934}. The interaction with the ground-state atom hybridizes the degenerate $\lvert n,\circlearrowright\rangle$ and $\lvert n,\circlearrowleft\rangle$ states into an equal-weight superposition which generates strong confinement for the atom along the radial and polar directions provided $a_s<0$, as is the case in Sr ($a_s = -13.2\,a_0$) \cite{DeSalvo2015}. The quantized quasi-harmonic atomic motion in these two directions can reach oscillation frequencies of a few MHz even for large $n$ due to the radial localization of the electron. A radial cut through the Born-Oppenheimer potential energy surface for a single atom, together with the first three bound states, is shown in Fig.\ref{Fig1}(c). As the  interaction aims to maximize the electron density at the atom's position, the phase of the oscillating electronic density is locked to the latter and rotates with the atom along the azimuthal direction. Therefore, there exists no confinement in the azimuthal degree of freedom and the entire dimer rotates as a rigid rotor.

Realizing an electronic lattice potential which governs the azimuthal dynamics thus requires at least two ground-state atoms inside the circular Rydberg orbital. This triatomic system, with the Rydberg atom trapped (e.g. via an optical tweezer) at the origin of the coordinate system, is described by the trimer Hamiltonian
\begin{equation}
 \hat H^{(3)}=
 \sum_{j=1}^{2} \left(-\frac{\hbar^2}{2 m_a}\nabla_{\bm R_j}^2
 +V_{\mathrm{ea}}(\bm r,\bm R_j)\right) \, ,
 \label{eq:microscopic-trimer}
\end{equation}  
where $m_a$ is the Sr atomic mass. We make use of the fact that the azimuthal atomic motion is slow compared to the fast radial ($\rho$) and axial ($z$) oscillations, to derive an effective low-energy angular Hamiltonian by averaging Eq.~\ref{eq:microscopic-trimer} over the diatomic vibrational ground-state density in $\rho$ and $z$  and projecting it into the degenerate circular state subspace $\{\lvert n,\circlearrowright\rangle, \lvert n,\circlearrowleft\rangle\}$ (see End Matter).
Up to a constant energy offset, the effective Hamiltonian reads
\begin{align}
 \hat H_{\mathrm{ang}}=\sum_{j=1}^2\left[-J\frac{\partial^2}{\partial \varphi_j^2}\identity+V\begin{pmatrix}
 0&e^{2i\ell\varphi_j}\\
 e^{-2i\ell\varphi_j}&0
 \end{pmatrix}\right].
 \label{eq:trimer-angular}
\end{align}
The parameters $V$ and $J$ set, respectively, the electronic lattice depth and the rotational constant of a single atom confined inside one of the circular orbitals. They are determined by averaging the Born-Oppenheimer surface $\langle n,\circlearrowright \lvert V_{\mathrm{ea}}({\bf r},{\bf R_q}) \lvert n,\circlearrowright\rangle$ and the rotational constant $\hbar^2/(2 m_a\rho_j^2)$ over the ground-state dimer density profile (see End Matter). $V$ and $J$ are both influenced by the choice of atomic species: $V$ is proportional to the electron-atom scattering length while $J$ is inversely proportional to the atomic mass.

From the behavior of the circular state wave functions at large $n$, we obtain  $J/V\sim n\frac{m_e}{m_a}\frac{a_0}{a_s}$. For $n=80$, one finds $V/h= {-7.937} \,\mathrm{MHz}$ and $J/h=0.513\,\mathrm{kHz}$. The off-diagonal potential in $\hat H_{\mathrm{ang}}$ couples $\lvert n,\circlearrowright\rangle$ to $\ket{n,\circlearrowleft}$ with a strength periodic in the atomic positions, as described above for the single-atom case. Now, however, both atoms compete to maximize the electron density at their position, giving rise to an effective long-range interaction which manifests itself as an electronic lattice potential governing their relative motion.

To see this, we transform into center-of-mass ($\Phi=(\varphi_1+\varphi_2)/2$) and relative ($\varphi=\varphi_1-\varphi_2$) azimuthal coordinates. An additional basis change into the electronic superposition states $\lvert n,\pm\rangle=1/\sqrt{2}(e^{i\ell\Phi}\lvert n,\circlearrowright\rangle \pm e^{-i\ell\Phi}\ket{ n,\circlearrowleft})$, i.e. standing waves with orientation determined by $\Phi$, removes the \(\Phi\)-dependent phases from the interaction. The Hamiltonian $\hat H_\mathrm{ang}$ then commutes with the center-of-mass momentum operator $\hat p_\Phi$. Consequently, the trimer states in the azimuthal coordinates, 
\begin{equation}
\Psi_K(\Phi,\varphi,\vec r)=\sum_{\alpha\in\{+,-\}} \left[e^{iK\Phi}w_{K,\alpha}(\varphi)\right]\langle \vec r|n,\alpha\rangle,
\end{equation}
are simultaneous eigenfunctions of $\hat p_\Phi$ and the Hamiltonian 
\begin{equation}
 \hat H_K = \left[\frac{JK^2}{2}
 -2J\partial_\varphi^2\right]\identity +J\ell K\hat\sigma_x +2V\cos(\ell\varphi)\hat\sigma_z,
 \label{eq:HK}
\end{equation}
where \(K\), the angular momentum of the center of mass, is a conserved quantum number. 

\begin{figure}[t]
  \centering
  \includegraphics[width=\linewidth]{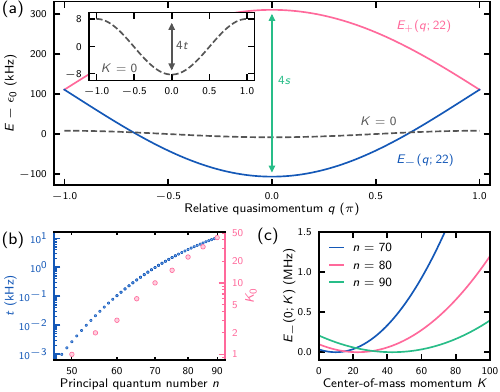}
  \caption{Band structure dictating the relative motion of two atoms in the electron lattice. (a) Lowest-band energy dispersions $E_\pm(q;K)$ for center-of-mass momenta $K=0$ and $K=22$. The inset shows a zoom into the $K=0$ band. Arrows indicate the $K=0$ bandwidth $4t$ and the $K=22$ band splitting $4s = 4(J\ell K)g$, respectively. (b) Intralattice tunneling parameter $t$ and center-of-mass angular momentum $K_0$ at which the trimer system has its lowest band energy as a function of principal quantum number $n$. $K_0$ sets the effective mass of interlattice tunneling dynamics via the interband tunneling $J\ell K_0g$. (c) Lowest band energy $E_-(0;K)$ as a function of $K$ for exemplary values of $n$ as indicated. For better comparison, the $n$-dependent offset $ E_-(0;K_0)$ has been subtracted. }
  \label{Fig2}
\end{figure}

The off-diagonal coupling between the electronic orbital angular momentum and the center-of-mass angular momentum acts as an effective spin-orbit coupling \cite{Galitski2013,Geier2023}. For $K=0$, this coupling vanishes and the Schrödinger equation reduces to two uncoupled systems with the Born-Oppenheimer potentials $U_\pm(\varphi)=\pm2V\cos(\ell\varphi)$, which form a pair of interlaced sinusoidal sublattices displaced by $\pi/\ell$ in the relative azimuthal angle (Fig.~\ref{Fig1}(d)). The center-of-mass motion ($|K|>0$), or equivalently, Larmor precession induced by an external $B$-field (see End Matter), couples the two sublattices via the $\hat \sigma_x$-term in $\hat H_K$. This allows us to identify the two processes governing the dynamics: tunneling within each sublattice (green arrow in Fig.~\ref{Fig1}(d)) and non-adiabatic coupling leading to tunneling between the sublattices (orange arrow in Fig.~\ref{Fig1}(d)). 

To derive the respective tunneling rates, we turn to a momentum-space representation. In the deep-lattice limit, projection onto the Wannier orbital of the lowest band yields the Bloch Hamiltonian
\begin{equation}
 H(q)=\begin{pmatrix}
 \epsilon_K-2t\cos q & J\ell Kg(1+e^{-i q})\\
 J\ell Kg(1+e^{i q}) & \epsilon_K-2t\cos q
 \end{pmatrix},
 \label{eq:bloch}
\end{equation}
where $\epsilon_K$ is the on-site energy, $t$ is the tunneling amplitude between the neighboring sites of a single sublattice $U_{\pm}$, and $J\ell Kg$ (simply referred to as $s$ in Fig.~\ref{Fig1}(d)) is the tunneling amplitude \textit{between} the sublattices, determined by the dimensionless overlap $g$ between neighboring Wannier orbitals of different lattice potentials (\textit{c.f.} Fig.~\ref{Fig1}(d)). Finally, $q=2 \pi k /\ell \in[-\pi,\pi)$ is the dimensionless quasimomentum conjugate to the unit-cell with lattice spacing $2\pi/\ell$ ($k$ is the angular quasimomentum). Diagonalizing Eq.~\ref{eq:bloch} yields the $K$-dependent dispersion
\begin{equation}
 E_\pm(q;K)=\epsilon_K-2t\cos q
 \pm\left|2J\ell Kg\cos\frac q2\right|.
 \label{eq:bands}
\end{equation}

Figure~\ref{Fig2}(a) illustrates how the system parameters shape the band structure. For $K=0$, tunneling between the sublattices vanishes. This results in a degenerate band doublet, corresponding to the two sinusoidal sublattices, with a cosine dispersion and a comparatively narrow bandwidth $4t$ (inset in Fig.~\ref{Fig2}(a)). By fitting the lowest band of the numerically calculated spectrum of $\hat H_{K=0}$ to Eq.~\ref{eq:bands}, we obtain the tunneling amplitudes $t$ as a function of $n$ (Fig.~\ref{Fig2}(b)). These show that appreciable motion around the electronic ring lattice requires large $n$. In particular, for $n=80$, we obtain $t/h=4.10\,\mathrm{kHz}$. Repeating this fitting procedure for finite $K$ yields $s/h=K\times 4.17\,\mathrm{kHz}$, corresponding to $g\sim0.1$, and $\epsilon_K\approx a\frac{JK^2}{2}+2V + \ell\sqrt{2J|V|}$. The factor $a\approx 0.75$ shows that the dispersion is modified by coupling to higher Wannier orbitals via the off-diagonal potential $J\ell K\hat \sigma_x$; if this term is neglected we obtain $a = 1$. This coupling additionally lifts the degeneracy between the two bands everywhere except at the Brillouin-zone edges $q=\pm\pi$, where the form factor $1+e^{-iq}$ vanishes and the two bands meet.

The $K$-dependent coupling has an important consequence. The band minimum $E_-(0; K)$ first decreases with increasing $K$, as the system may lower its energy via interlattice tunneling. However, competition with the increasing rotational energy of the center of mass, $\propto K^2$, leads to a minimal system energy $E_-(0;K_0)$, where $K_0\approx \frac{2\ell g}{a}$ depends strongly on $n$ (Fig.~\ref{Fig2}(c)). Note that $K_0$ also depends on the strength of the electron-atom scattering length through $g/a$, as the Wannier orbital localization increases with larger $a_s$. For our example principal quantum number, $n=80$, one obtains $K_0\sim22$, and $K_0$ derived for a range of $n$-values is shown in Fig.~\ref{Fig2}(b). Over this range, $K_0$ scales algebraically with $n$. As we will see below, the value of $K_0$ sets the effective mass of the atoms tunneling through the electronic lattice and, consequently, governs the two-particle dynamics that we study next.

To this end, we propose a scheme to probe the tunneling motion experimentally. Importantly, the strong localization of the circular Rydberg electron yields deep lattices and strong radial confinement even at comparatively large $n$ where the orbital diameter approaches the micrometer scale. This enables local control of the atoms within the induced electronic lattice using optical tweezers. We consider a situation where the Rydberg atom and the two ground-state atoms are initially trapped in independent tweezers separated by distances much larger than the Rydberg orbit. In the tweezer ground state, the relative position of the atoms can be controlled at the $10$ nm level \cite{Guttridge2025}, which is comparable to the radial width of the circular electron orbit ($\approx 40 \, \mathrm{nm}$ for $n=80$). Using species selective trapping conditions, e.g. by choosing a tune-out wavelength \cite{Heinz2020}, the atoms can then be radially moved into the circular orbit to end up adiabatically in the ground state of the summed tweezer and electron potential, similar to recent demonstrations of mergo-association of Feshbach dimers \cite{Ruttley2023,Stock2003}. Overlap with the Rydberg electron may be probed by autoionization upon optical excitation of the ground state, while quantized radial bound states are accessible via microwave spectroscopy between adjacent circular states.

\begin{figure}[t]
  \centering
  \includegraphics[scale=1]{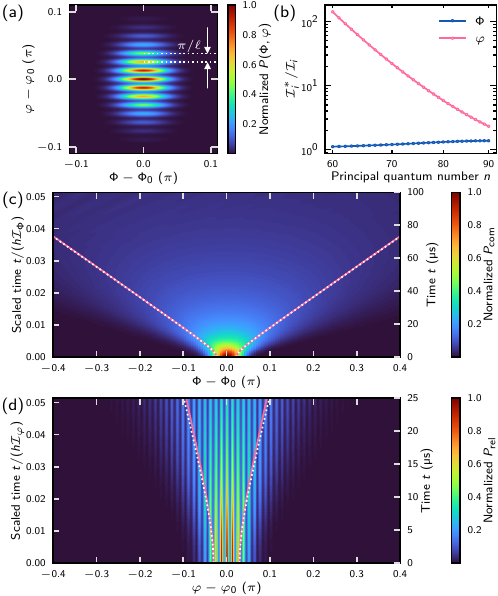}
  \caption{Two-atom tunneling dynamics in the electronic lattice. (a) Density of the initial state in the combined circular state and tweezer potential after integrating over the electronic degree of freedom for $\kappa/h=2.06\,$MHz, $\Phi_0=\pi$, and $\varphi_0=-\pi$. (b) Ratio of the effective moment of inertia for center-of-mass (blue) and relative (pink) motion in the electronic lattice to the bare moment of inertia without the lattice. (c,d) Time evolution of the two-atom correlators $P_{\rm{com}}$ and $P_{\rm{rel}}$ in the center-of-mass and relative motion after a sudden removal of the tweezer potential $U_{\mathrm{tw}}$. The solid line indicates the fitted rms-width of the numerical data, whereas the dashed line shows the width $\sigma_i(t)$ for ballistic expansion with the effective moments of inertia $\mathcal I_i^*$.}
  \label{fig:dynamics}
\end{figure}

After this process, the atoms are adiabatically prepared in the ground state of the Hamiltonian $\hat H_\mathrm{init}=\hat H_\mathrm{ang} +U_\mathrm{tw}$, where
\begin{equation}
 U_{\mathrm{tw}}(\Phi,\varphi)=\frac{\kappa}{4}
 \left[4(\Phi-\Phi_0)^2+(\varphi-\varphi_0)^2\right],
 \label{eq:tweezer}
\end{equation}
is the additional tweezer confinement with angular trapping strength $\kappa$ controlling the initial angular localization, and $\Phi_0 (\varphi_0)$ denotes the tweezer center-of-mass (relative) coordinate at which the atoms are trapped. The two-atom ground state is shown in Fig.~\ref{fig:dynamics}(a). Its density displays both a Gaussian envelope imposed by the tweezer confinement and a lattice-induced density modulation along the relative coordinate \(\varphi\). The tweezers holding the two atoms are then rapidly switched off, allowing this initial state to expand in the lattice. The numerically calculated time evolution after this quench is shown in Fig.~\ref{fig:dynamics}{{(c,d)}}. Specifically, we display the two-atom dynamics in the center-of-mass and the relative coordinate, obtained via $P_\mathrm{com}(\Phi,t) = \int d\varphi|\Psi(\Phi,\varphi,t)|^2$ and $P_\mathrm{rel}(\varphi,t) = \int d\Phi| \Psi(\Phi,\varphi,t)|^2$, respectively. Both types of dynamics are directly accessible experimentally by either extracting single-atom positions or measuring atom-pair correlations via recapture and spatial readout with tweezers.

The results reveal a clear ballistic expansion in both $\Phi$ and $\varphi$, but at different rates, determined by the lattice's role in defining the effective moment of inertia. To see this, we indicate the rms-widths of the numerical data (solid lines) and compare them with a ballistic expansion model in which the wavepacket width evolves as $\sigma_i(t)= \sqrt{\sigma_{i}^2(0) + \left(\sigma_{p_i}t/\mathcal I_i^*\right)^2}$ (dashed lines). Here,  $\sigma_{p_i}$ is the initial angular-momentum width, $\mathcal I_i^*$ is the effective moment of inertia, and $i\in\{\Phi,\varphi\}$. As there is no periodic potential along the $\Phi$ coordinate, the spreading of $P_{\rm{com}}$ is close to what one would expect from a free rotor with bare moment of inertia $\mathcal I_\Phi=1/J$. However, coupling between the center-of-mass rotation and the electronic doublet leads to a small modification. Our numerical data is found in excellent agreement with an effective moment of inertia $\mathcal I_\Phi^* \simeq \mathcal I_\Phi/a$, which one obtains from the curvature of $E{_-}(0;K)$ at $K_0$ (see End matter). 

In contrast, the two-atom dynamics in the relative coordinate $P_{\rm{rel}}$ evolves within the electronic lattice, as is strikingly illustrated by the pronounced density maxima visible at each lattice site in Fig.~\ref{fig:dynamics}(d). Tunneling through the lattice notably slows down the expansion compared to what one would expect from the bare moment of inertia in the $\varphi$ coordinate, $\mathcal I_\varphi=1/(4J)$, without the lattice. The numerical data is found in good agreement with the effective inertia $\mathcal I_\varphi^* = 5.8 \mathcal I_\varphi$, determined from the curvature of the tight-binding band $E{_-}(q,K_0)$ at $q=0$. Note that, although the initial state contains contributions from many $K$ values (see Fig.~\ref{fig:Inital_state} in the End Matter), its dynamics are dominated by $K$ values near its peak value $K_0$. The interlattice coupling and associated band curvature for $K=K_0$ therefore set the characteristic scale of the expansion in $\varphi$ through the effective moment of inertia. In Fig.~\ref{fig:dynamics}(b), we plot the $n$-dependence of $\mathcal I_i^*$, showing once more that the electronic lattice effects the relative motion of the atoms, and yields a strong increase of $\mathcal I_i^*$ with decreasing $n$.

To summarize, the electronic lattice manifests as a genuine two-particle effect. The tunneling dynamics are revealed in the atom-pair correlations, while the individual atoms expand essentially as a free rotors. The latter determines the tunneling rate in the electronic lattice through the strength of the motion-induced coupling between the circular orbitals that hybridize to form the electronic standing wave. For three atoms we anticipate strong genuine three-body interactions, which will be subject of future work.

In conclusion, we have shown that long-lived electronic Rydberg orbitals give rise to micrometer-scale potentials for ultracold atoms, and specifically lattice dynamics with intrinsic coupling between atomic and electronic degrees of freedom. These systems can be readily realized combining recent advances in Rydberg state control and neutral atom tweezer technology. Our work opens routes to explore additional Hubbard-like interactions between the atoms, mass-imbalanced trimers, and trapping of more than two atoms inside circular electronic potentials. Perspectively, we envision trapping even small, interacting Bose or Fermi gases within the annular Rydberg potential surface and exploring the role of quantum statistics and correlations in their dynamics. The strong and long-range two-body interaction induced by the circular state electron calls for a generalization to many-body interactions in future work.

\textit{Acknowledgments:} F. M. acknowledges funding from the Federal Ministry of Research, Technology and Space under the Grant CiRQus. T. P. acknowledges funding from the European Research Council (ERC)(Grant Agreement No. 101019739). F.M. and T.P. acknowledge funding from the Federal Ministry of Research, Technology and Space under the Grant QRydDemo, the Horizon Europe Programme HORIZON-CL4-2021- DIGITAL-EMERGING-01-30 via Project No. 101070144 (EuRyQa), and the Carl Zeiss Foundation via IQST. 

\bibliographystyle{apsrev4-2}

%

\newpage
\section{End Matter}

\subsection{Trimer Hamiltonian, $B$ field, and initial state}
In an electric field configuration where the circular states $\ket{n\circlearrowright}$ and $\ket{n\circlearrowleft}$ are energetically isolated from the rest of the Rydberg manifold, the full vibronic wave function can be expanded as
\begin{align}
    \Psi(\vec R_1,\vec R_2,\vec r) &= \sum_{\circ=\circlearrowleft}^{\circ=\circlearrowright}\chi_\circ(\vec R_1,\vec R_2)\langle\vec r\ket{n,\circ}.
\end{align}
Applying $\hat H^{(3)}$ (Eq.~\ref{eq:microscopic-trimer}) to $\Psi$ and then integrating over the electronic states leads to the two-level Hamiltonian  
\begin{align}
    \hat H^{(3)} &= \sum_{j=1}^2\left[(\hat T_j + D_{jn})\identity + D_{jn}\begin{pmatrix}
        0 & e^{-2i\ell\varphi_j} \\e^{2i\ell\varphi_j} & 0
    \end{pmatrix}\right],\nonumber
\end{align}
where $D_{jn}\equiv D_n(\rho_j,z_j) = \bra{n,\circlearrowright}V_{ea}\ket{n,\circlearrowright}$ is independent of $\varphi_j$ and of the circulation direction $\circlearrowright$. 
The first term of $\hat H^{(3)}$ is the sum of two diatomic Hamiltonians and two azimuthal rotational kinetic energies,
\begin{align}
    \sum_{j=1}^2(\hat T_j + D_{jn})\identity  &= \left(\sum_{j=1}^2 H_j^{(2)}-\sum_{j=1}^2\frac{1}{2 m_a\rho_j^2}\frac{\partial^2}{\partial \varphi_j^2}\right)\identity,
\end{align}
where $H_j^{(2)}=\left(\hat T^0_j + D_{jn}\right)$ is the Hamiltonian describing the vibrational dynamics of a single ground-state atom interacting with the Rydberg atom. 
The trimer wave function in the electronic state $\ket{n,\circ}$ can be expanded into all stationary states of the two independent dimers:
\begin{align}
    \chi_\circ(\vec R_1,\vec R_2)=\sum_{\nu,\mu}\chi_{\nu}(\rho_1,z_1)\chi_\mu(\rho_2,z_2)W_{\nu,\mu,\circ}(\varphi_1,\varphi_2).\label{eq:ansatz}\nonumber
\end{align}
Due to the comparatively large separation between vibrational dimer energies (see Fig.~\ref{Fig1}(c)) and the fact that the matrix elements $V_{\mu\nu}=\int \chi_\mu^*(\rho_j,z_j)D_{jn}\chi_{\nu}(\rho_j,z_j)\rho_jd\rho_jdz_j$ nearly vanish for odd $\mu + \nu$, it is a good approximation to consider only the $\mu = \nu = 0$ contribution. We therefore average over the vibrational ground state, defining
\begin{equation}
    V \equiv V_{00},\,
    \varepsilon_0 \equiv
    \left\langle \chi_0 \middle| H_j^{(2)} \middle| \chi_0 \right\rangle,\,
    J \equiv
    \left\langle \chi_0 \middle| \frac{1}{2\mu\rho^2} \middle| \chi_0 \right\rangle,\nonumber
\end{equation}
to isolate the Hamiltonian governing the lattice dynamics,
\begin{align}
    \hat H_\mathrm{ang} &=\sum_{j=1}^2\left[(\varepsilon_0 -J\partial_{\varphi_j}^2 )\identity+V\begin{pmatrix}
        0 & e^{-2i\ell\varphi_j} \\  e^{2i\ell\varphi_j} & 0
    \end{pmatrix}\right] .\nonumber
\end{align}
We perform a change of variables to the center-of-mass ($\Phi =\frac{\varphi_1 + \varphi_2}{2}$) and relative ($\varphi = \varphi_1 - \varphi_2$) coordinates, in which the Hamiltonian reads
\begin{align}
\begin{split}
    \hat H_\mathrm{ang} =& \left[2\varepsilon_0 - \frac{J}{2}\left(\frac{\partial^2}{\partial\Phi^2} +4\frac{\partial^2}{\partial\varphi^2}\right)\right]\identity\\&+2V\begin{pmatrix}0 & e^{-2i\ell\Phi}\cos(\ell\varphi)\\e^{2i\ell\Phi}\cos(\ell\varphi) & 0
    \end{pmatrix}. 
    \end{split}
\end{align}
We diagonalize the potential term by applying 
\begin{align}
   \hat U &= \frac{1}{\sqrt{2}}\begin{pmatrix}
        -e^{i\ell\Phi} & e^{-i\ell\Phi}\\
        e^{i\ell\Phi} & e^{-i\ell\Phi}
    \end{pmatrix}
\end{align}
to  $\hat H_\mathrm{ang}$, leading to
\begin{align}
    \tilde H_\mathrm{ang} =& \left[ \frac{J}{2}\left(p_\Phi^2 + 4p_\varphi^2\right) \right]\identity- 2V\cos(\ell\varphi)\hat\sigma_z  +J\ell\hat p_\Phi\hat\sigma_x,\nonumber
\end{align}
where $\hat p_x = \frac{1}{i}\frac{\partial}{\partial x}$ and the overall energy offset $\Delta = 2\varepsilon_0 + \frac{J\ell^2}{2}$ is set to zero. 
$\tilde H_\mathrm{ang}$ commutes with $\hat p_\Phi$ and thus 
\begin{align}
    \Psi_K(\Phi,\varphi,\vec r) =  e^{iK\Phi}\sum_{\substack{\alpha\in\{+,-\}\\\circ\in\{\circlearrowright,\circlearrowleft\}}} w_{K,\alpha}(\varphi)U_{\alpha\circ}\langle \vec r | n,\circ\rangle,
    \label{eq:chi_separation}
\end{align}
where $K$ is the total angular momentum, is a simultaneous eigenfunction of $\tilde H_\mathrm{ang}$ and $\hat p_\Phi$. The function $w_{K,\circ}(\varphi)$ is obtained by diagonalizing $\hat H_K$ (Eq. \ref{eq:HK}).

Adding a magnetic field $B$ (along $z$) lifts the degeneracy between the circular states $\ket{n\circlearrowright}$ and $\ket{n\circlearrowleft}$ by the Zeeman shift {$\Delta_B=\mu_B \ell B$}, where $\mu_B$ in the Bohr magneton. When adding this to the trimer Hamiltonian, the off-diagonal coupling term in $\hat H_K$ becomes $(JlK - \Delta_B) \hat \sigma_x$. This is a manifestation of Larmor precession of the electronic standing wave \cite{Dietsche2019}, which plays an equivalent role as finite $K$. We may thus estimate the required cancellation of magnetic fields to $B \ll JK/\mu_\text{B} \approx 8.1~\text{mG}$ (for $K=22$ and $n=80$) to experimentally observe the lattice dynamics described in the main text. 

The initial state is obtained by adding the tweezer potential $U_\mathrm{tw}(\Phi,\varphi)$ (Eq.~\eqref{eq:tweezer}) to $\hat H_\mathrm{ang}$ and then solving for the ground state of the resulting Hamiltonian. As the tweezer potential breaks both the lattice symmetry and the rotational symmetry, the initial state is a superposition of angular momentum eigenfunctions $e^{ik\varphi}$ and $e^{iK\Phi}$ in both the relative and center-of-mass coordinates. The distributions of these momenta in the initial state are shown in Fig.~\ref{fig:Inital_state}. The $P_K$ distribution is peaked at $K_0$ as discussed in the main text (only $K>0$ is shown here), while the relative-momentum distribution \(P_k\) exhibits a central peak and additional Bragg peaks at \(k=\pm\ell\).

\subsection{Tight-binding parameters}
The dimensionless parameter $d=\frac{8|V|}{J\ell^2}$ characterizes the depth of the lattice.  
In a deep lattice, $d\gg1$, the intralattice tunneling rate $t$ is approximately \cite{bloch2008many}
\begin{equation}
 t\simeq\frac{2J\ell^2}{\sqrt\pi}d^{3/4}e^{-2\sqrt{d}}.
 \label{eq:t-wkb}
\end{equation}
The interlattice tunneling amplitude $s=J\ell Kg$
is similarly determined by the overlap $g$ between Wannier orbitals belonging to the two shifted lattices. As this property is especially sensitive to the long-range tails of neighboring Wannier orbitals, we extract $g$ (and similarly, $a$) numerically from fitting the spectrum to Eq.~\ref{eq:bands}. 

\subsection{Effective moments of inertia}

The effective moment of inertia of the center-of-mass coordinate is obtained from the curvature of the lowest eigenenergy $E_-(0;K)$ near its minimum at $K = K_0$. We fit the numerical spectrum locally to
\begin{equation}
 E_-(0;K)\simeq E_-(0;K_0)+\frac{aJ}{2}(K-K_0)^2.
\end{equation}
Comparing with the free dispersion $(K-K_0)^2/(2\mathcal I_\Phi^*)$ gives
$\mathcal I_\Phi^*/\mathcal I_\Phi =1/a\simeq1.3$ for $n=80$.  

\begin{figure}
    \centering
    \includegraphics[width=\linewidth]{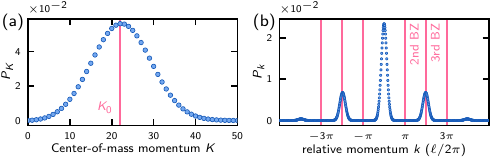}
    \caption{Center-of-mass and relative angular momentum distributions, $P_K$ and $P_k$, of the initial state confined by the tweezers. (a) The ground-state composition is determined by $K_0$, resulting in a center-of-mass momentum distribution $P_K$ peaked at this value and symmetric about $K=0$. (b) The Bragg peaks in $P_k$ result from the density modulation with lattice constant $\pi/\ell$ imposed onto the initial density. Consequently, the first order Bragg peak sits at the edge of the 2nd Brillouin zone of the $2\pi/\ell$ periodic sublattices.}
    \label{fig:Inital_state}
\end{figure}

For relative motion at fixed $K$, expansion of the lower tight-binding band around $k=0$ yields 
\begin{equation}
 E_{-}(k;K)\simeq E_{-}(0;K)
 +\frac{2\pi^2}{\ell^2}\left(2t+\frac{|J\ell Kg|}{2}\right)k^2,
\end{equation}
from which we obtain
\begin{equation}
 \mathcal I_\varphi^*(K)
 =\frac{(\ell/\pi)^2}
 {8t+4|J\ell Kg|}.
\end{equation}
This is a good approximation for states that are sufficiently well-localized about the band minimum. 
Our numerical data (Fig.~\ref{fig:dynamics}) shows a slight deviation from this quadratic approximation, which we attribute to the width of the initial state in the first Brillouin zone (c.f.  Fig.~\ref{fig:Inital_state}(b)). 

\subsection{Effective two-state Rydberg electron model}

Finally, we discuss the feasibility of the effective two-state Rydberg electron basis used here. The hydrogenic manifold of high angular-momentum Rydberg states has a large number of $n^2$ degenerate substates. Applying an external electric field $F$ lifts this degeneracy via the Stark effect. The hydrogenic eigenstates are then described in the parabolic basis~\cite{Stebbings2011} with the electronic quantum number $k$ and are shifted up to second order by
\begin{align}\label{eq:Stark_shifts}
    \Delta^{n}_{k,m} = \frac{3}{2}nkF - \frac{n^4}{16}(17n^2 - 3k^2 - 9m^2 + 19)F^2.
\end{align}
The first term describes the linear Stark shift, directly proportional to the intrinsic dipole moment determined by $k$. This allows us to separate the circular states with $\Delta^n_{0,\ell}$ from the $k\neq0$ states at relatively low electric fields. The second order contribution in Eq.~\eqref{eq:Stark_shifts} lifts the remaining degeneracy in $k$ via $|m|$-dependent second-order shifts. To mitigate avoided crossings throughout the lattice potential, the relative red-shift $\Delta^n_{0, \ell-2} - \Delta^n_{0,\ell}$ of the closest non-circular states needs to exceed the potential depth $4V$. In our setup, this requires electric fields of $F \gtrsim \SI{4.2}{\volt\per\centi\meter}$. At such fields highly polarized Rydberg states from neighboring $n$ manifolds cross over $\Delta^n_{0,\ell}$ and need to be evaded. We neglect polarization of the circular wavefunction, which introduces an offset in $z$.

\begin{figure}
    \centering
    \includegraphics[width=1\linewidth]{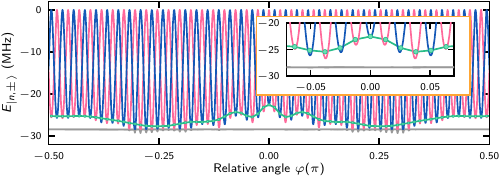}
    \caption{Electronic lattice including admixture of hydrogenic states at an electric field $F'$. Perturbed eigenenergies of electronic lattice states $\ket{n, +}$ (blue) and $\ket{n, -}$ (pink) have minima modulated with a period of $\pi/2$ along $\varphi$. Integration over a vibrational ground state at multiples of $\pi/l$ results in reduced site-to-site fluctuations (see inset) and slightly reduced well depths. Avoided crossings with $\ket{n,k=0,|m|=n-3}$ (gray) can be minimized.}
    \label{fig:perturbed_lattice}
\end{figure}

With the above considerations, we find suitable settings at $F'=\SI{3.97}{\volt\per\centi\meter}$. We investigate the perturbations on the lattice potential by diagonalizing $V_{ea} + \Delta^n_{k,m} -\Delta^n_{0,\ell}$ for a basis spanning all $m$ states for $k\in[-1,1]$ in $n$ and $n' \neq n$ states with $k$ crossing within a $\SI{0.2}{\volt\per\centi\meter}$ range around $F'$. The resulting eigenenergies $E_{\ket{n,\pm}}$ of states most closely resembling the electronic lattices $\ket{n,\pm}$ are shown in Fig.~\ref{fig:perturbed_lattice}. We find slow modulations on the lattice potential minima which are dominated by admixture of non-circular states with $k=0, |m|=\ell-4$, resulting in a period of $\pi/2$, whereas nearby $n'\neq n$ states mainly reduce the lattice depth and thereby help avoid the $\ket{n,0,\ell-2}$ state. Integrating out the $(\rho, z)$-dependence by assuming a vibrational ground state of the dimer potential as shown in Fig.~\ref{Fig1}(b), the site-to-site depth fluctuations are further reduced.

\clearpage
\newpage

\begin{widetext}
\section{Supplementary Material}
This supplement contains some additional derivations and details helpful for understanding the $n$-scaling of various quantities in the main text and presents additional band gap calculations. In the supplement, we work in atomic units. 

The Rydberg atom is prepared in the circular state denoted as $\ket{n \, \circlearrowright} \equiv \ket{n,\ell=n-1,m= (n-1)}$ or  $\ket{n \, \circlearrowleft} \equiv \ket{n,\ell=n-1,m= -(n-1)}$.    The circular state (using $\circ$ as a generic label for either $\circlearrowright$ or $\circlearrowleft$) has the position-space wave function \begin{equation}
    \Psi_{n\circ}(\vec r) = R_{n\ell}(r)Y_{\ell\circ}(\hat r),
\end{equation} where
    \begin{align}
        R_{n\ell}(r) &= \frac{2^n}{n^2\sqrt{(2n-1)!}}\left(\frac{r}{n}\right)^{n-1}e^{-r/n}\\
        Y_{\ell\circ}(\hat r) &= \frac{(-1)^{n-1}}{2^{n-1}}\sqrt{\frac{(2n-1)!}{4\pi}}\frac{\sin^{n-1}(\theta_r)}{(n-1)!}\left(e^{i\ell\varphi_r}\delta_{\circ\circlearrowright}+e^{-i\ell\varphi_r}\delta_{\circ\circlearrowleft}\right).
\end{align}
Assuming just $s$-wave scattering with a constant scattering length $a_s$, the pseudopotential is 
    \begin{equation}
        V_{\mathrm{ea}}(\vec r,\vec R) = 2\pi a_s\delta^3(\vec r - \vec R).
    \end{equation}
In the two-level electronic basis $\{|n,\circlearrowleft\rangle,\ket{n,\circlearrowright}\}$ two matrix elements are relevant: 
\begin{align}
    \bra{n\circlearrowleft}V_{\mathrm{ea}}\ket{n\circlearrowleft} = \bra{n\circlearrowright}V_{\mathrm{ea}}\ket{n\circlearrowright}&= \frac{2a_s}{n^4[(n-1)!]^2}\frac{\rho^{2(n-1)}}{n^{2(n-1)}}e^{-2\sqrt{\rho^2+z^2}/n}\\
    \bra{n\circlearrowright}V_{\mathrm{ea}}\ket{n\circlearrowleft}=[\bra{n\circlearrowleft}V_{\mathrm{ea}}\ket{n\circlearrowright}]^* &= \frac{2a_s}{n^4[(n-1)!]^2}\frac{\rho^{2(n-1)}}{n^{2(n-1)}}e^{-2\sqrt{\rho^2+z^2}/n}e^{-2i(n-1)\varphi},
\end{align}
where we have integrated over the electronic coordinates and expressed the internuclear position in cylindrical coordinates. 
We define $D_{n}(\rho,z) = D_n(R,\theta)\equiv \bra{n\circlearrowleft}V_{\mathrm{ea}}\ket{n\circlearrowleft}$. 
Note that $D_n$ is independent of the direction $\circlearrowright$ or $\circlearrowleft$ and $\bra{n\circlearrowright}V_{\mathrm{ea}}\ket{n\circlearrowleft} = D_ne^{-2i(n-1)\varphi}$.  We can expand $D_n(\rho,z)$ to zeroth order about $\rho\sim n(n+1)$ and $z\sim 0$ to get the $n$-scaling of the lattice potential depth ($V$ in the main text)
\begin{align}
    D_n(\rho,z)&\approx 
    -\frac{2|a_s|e^{2-2n}(n-1)^{2n-3}}{n^6[(n-1)!]^2}n^2(n-1),
\end{align}
assuming that $a_s$ is negative here. Asymptotically, this goes as 

\begin{equation}
    D_n \sim \frac{2|a_s|}{2\pi}\frac{(n-1)^{-2}}{n^6}n^3\sim n^{-5},\quad n\to \infty
\end{equation}

Figure~\ref{fig:placeholder} shows the energy bands for three different values of the total center-of-mass angular momentum $K$ and for $n=80$. These highlight (a-c) the large bandgap between the lowest and first excited bands,  and (d-f) the dependence of the lowest band's width on $K$. 

\begin{figure}
    \centering
    \includegraphics[width=1\linewidth]{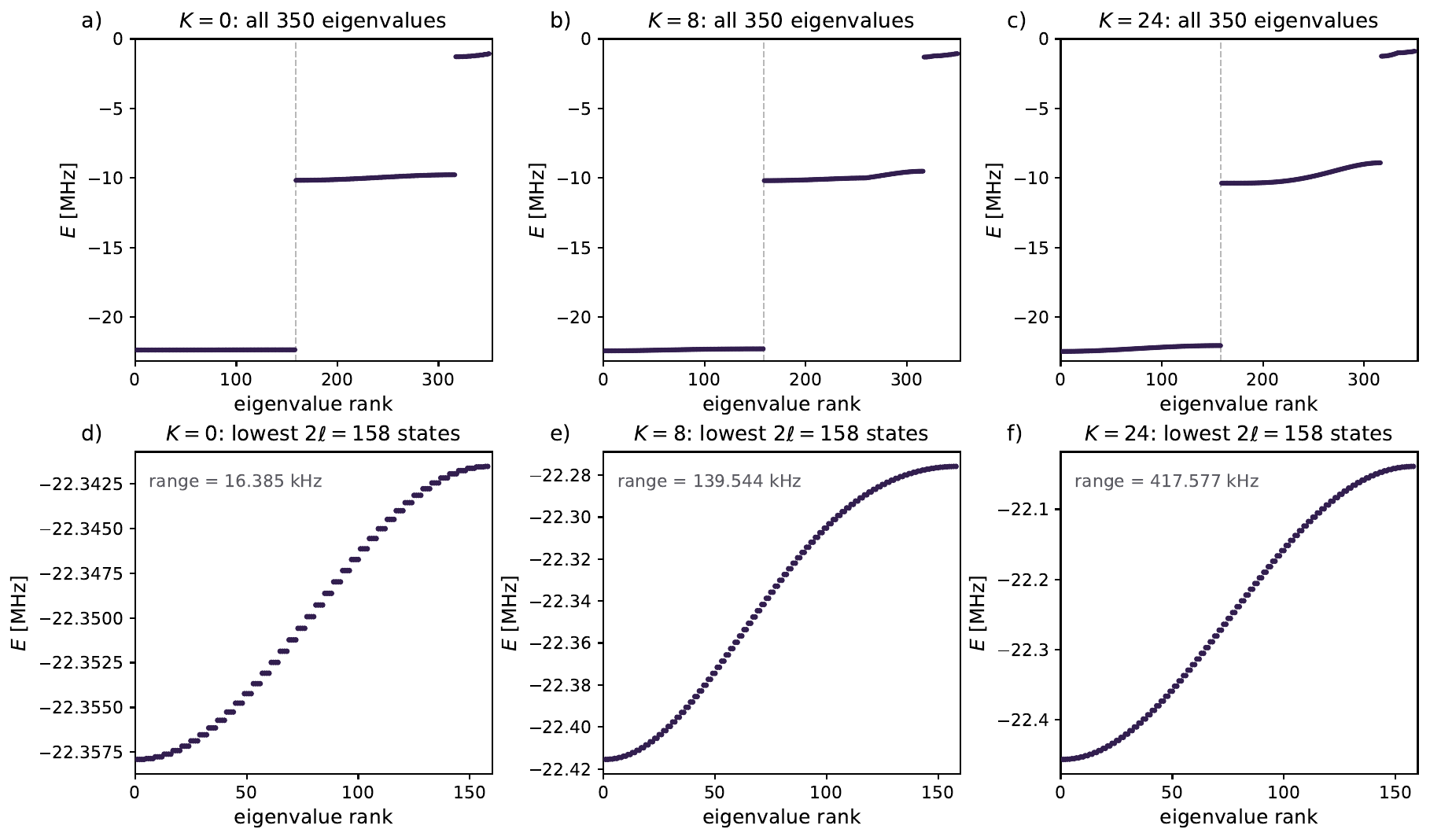}
    \caption{Numerically calculated circular-state Rydberg trimer spectra for three different values of the center-of-mass rotational angular momentum $K$. Panels (a–c) show the lowest 350 energy eigenvalues for \(n=80\) for  \(K=0\), \(K=8\), and \(K=24\), respectively. The vertical dashed line marks position of the first band gap; the lowest band contains \(2\ell=158\) states for \(\ell=n-1=79\). Panels (d–f) show the lowest band alone for the corresponding \(K\) values. The bandwidths for each $K$ value are \(16.385\,\mathrm{kHz}\), \(139.544\,\mathrm{kHz}\), and \(417.577\,\mathrm{kHz}\), respectively. The increasing width with \(K\) results from the spin-orbit coupling in $\hat H_K$.}
    \label{fig:placeholder}
\end{figure}

\end{widetext}

\end{document}